# Reconsidering Spatial Priors In EEG Source Estimation

## Does White Matter Contribute to EEG Rhythms?

Pamela K. Douglas
Computer Science, Modeling & Simulation, IST, UCF
Department of Psychiatry & Biobehavioral Medicine,
David Gefen School of Medicine, UCLA
pkdouglas16@ucla.edu

David B. Douglas
Nuclear Medicine & Molecular Imaging
Stanford Hospital
Palo Alto, CA
davidbdouglas@gmail.com

*Abstract* Electroencephalogram (EEG) has been a core tool used in functional neuroimaging in humans for nearly a hundred years. Because it is inexpensive, easy to implement, and noninvasive, it also represents an excellent candidate modality for use in the BCI setting. Nonetheless, a complete understanding of how EEG measurements (voltage fluctuations) relate to information processing in the brain remains somewhat elusive. A deeper understanding of the neuroanatomical underpinnings of the EEG signal may help explain inter-individual variability in evoked and induced potentials, which may improve BCI therapies targeted to the individual. According to classic biophysical models, EEG fluctuations are primarily a reflection of locally synchronized neuronal oscillations within the gray matter oriented approximately orthogonal to the scalp. In contrast, global models ignore local signals due to dendritic processing, and suggest that propagation delays due to white matter architecture are responsible for the EEG signal, and are capable of explaining the coherence between numerous rhythms (e.g., alpha) at spatially distinct areas of the scalp. Recently, combined local-global models suggest that the EEG signal may reflect a superposition of local processing along with global contributors including transduction along white matter tracts in the brain. Incorporating both local and global (e.g., white matter) priors into EEG source models may therefore improve source estimates. These models may also help disentangle which aspects of the EEG signal are predicted to colocalize spatially with measurements from functional MRI (fMRI). Here, we explore the possibility that white matter conductivity contributes to EEG measurements via a generative model based on classic axonal transduction models, and discuss its potential implications for source estimation.

I. INTRODUCTION

Human electroencephalogram (EEG) celebrates its 95th birthday this year (Berger 1929). Since its inception, EEG has provided unparalleled and non-invasive access to the temporal dynamics of the brain. EEG measures voltage fluctuations at the scalp that reflect the instantaneous superposition of electric dipoles in the brain, and numerous aspects of the EEG temporal signature have been mapped to cognitive tasks in healthy individuals and clinical populations alike (Lopez da Silva & Van Rotterdam 1987).

A. *Challenges with EEG for BCI*

EEG theoretically represents an excellent candidate modality for use in the brain computer interface (BCI) setting, however, many challenges for effective EEG-BCI implementation remain (Blankertz et al. 2006). For example, EEG research has a long history of characterizing transient time domain responses such as event related potentials (ERPs) (e.g., Woodman 2010). Classically, these studies employ signal averaging techniques to isolate responses over many trials. However, responses are known to vary considerably with repeated stimuli (Auksztulewicz et al. 2016) and across time within an individual (Müller, K.-R. Et al. 2008; Blankertz et al. 2006). Characterizing these non-stationarities remains a challenging issue for single-trial EEG decoding (Wojcikiewicz et al. 2011; Douglas et al. 2012).

Secondly, despite the long tenure and widespread use of EEG, certain aspects of the signal itself remain insufficiently characterized if not somewhat mysterious. For example, EEG power and frequency are inversely proportional on a logarithmic scale. This is a curious observation because many patterns found in nature (e.g., branching patterns in trees) also follow this $1/f^n$ power law behavior (Handel & Chung 1993). These robust (log-linear) rate dynamics appear to be scale invariant in the brain, preserved over time, and stable across a variety of perceptual contexts (Buzsáki 2006). Even when afferent information significantly alters individual (local) neural firing rates, the population firing rates are only modestly and transiently altered (Buzsáki & Mizuseki 2014). Predictive coding theory suggests that these mesoscopic neurodynamics may be energetically favorable, where Helmholtz free energy or cross-entropy between top-down brain activity and sensorial inputs from the environment acts as a form of cost function (Friston & Buszaki 2016). Although this theory represents an attractive organizing principle, it requires coupling with a biologically plausible generative model to describe how the brain's structure and function give rise to these patterns of oscillation (Bastos et al. 2012).

EEG spectral densities putatively reflect structural organization, network properties (Vecchio et al. 2017), or functional and effective connectivities (Steinke & Galan 2011). However, cross frequency phase-amplitude couplings are evident across the brain (Canolty 2010), and add another dimension of complexity to the EEG signal. For example, increases in gamma power coincide with the negative going phase of the delta rhythm in non-human primates during naturalistic visual processing (Whittingstall & Logothetis 2009), and detection of otherwise sub threshold visual stimuli appears to be mediated by interactions between ERP amplitudes and the phase of the alpha rhythm (Mattewson et al. 2009). Therefore, appreciating events and their variability within the context of oscillations is essential.

Unfortunately, EEG scalp recordings, or sensor space data, do not provide sufficient information to infer involvement between and amongst brain structures (Haufe et al. 2011; Mahjoory et al. 2017). Therefore, the utility of EEG measurements rests on solving a challenging question: *How can external EEG measurements at the scalp be used to determine the internal location in the brain responsible for generating the signal?* This problem requires solving both a forward model and then calculating inverse EEG source solutions. The latter inverse problem is considered "ill-posed" because there are an infinite combination of internal fluctuations that could propagate to the scalp to produce a particular measurement (Mahjoory et al. 2017). Answering this question is of vital importance for diseases such as epilepsy, where determining the precise location of a seizure onset zone is crucial for patients who require surgical resection (Hamandi et al. 2006; Cooray et al. 2016). A variety of solutions have been proposed for solving the EEG forward and inverse problems. These typically involve additional assumptions which may take the form of spatial constraints, penalty functions, or Bayesian estimators (e.g., Haufe 2011). These assumptions are critical and should be consistent with a neurobiologically plausible generative model of the EEG signal.

### B. A Unified Generative Model for EEG

Ideally, a unifying generative model of EEG should be able to capture the aforementioned dynamics and reproduce key characteristics of the EEG signal including: transient responses such as evoked potentials and their nonstationarities (Wojcikiewicz et al. 2011), discrete spectral peaks or rhythms (Robinson et al. 2001), $1/f^n$ power scaling (Buszaki et al. 2016), differences in log-dynamics observed during sleep (Toussaint et al. 2000), cross frequency and phase interactions (Canolty 2010), inter-individual differences such as the spectral position of the alpha peak (Valdez-Hernandez et al. 2009), and predict phenomena such as traveling neocortical waves (Zhang et al. 2018). This model should also be able to predict these dynamics and their variation based on morphological, geometric, and functional properties of the brain. Currently, it is unclear if such a unifying model exists.

By far, the most widely adopted approach is to assume that cortico-cortico interactions are responsible for generating EEG measurements, and that EEG sources reside either within the gray matter or at the boundary between gray and white matter (e.g., Lei et al. 2015). This assumption greatly reduces the solution space of possible source locations, making the localization problem more numerically tractable. This constraint is based on the widely held tenet that EEG primarily measures postsynaptic current densities associated with synchronous activity from neural ensembles within the neocortical sheet. From this perspective, dipole generators of the EEG signal must be approximately orthogonal to and proximal to the scalp (e.g., Scherg 1990).

### C. White Matter Propagation Model

In recent years, a number of papers have investigated aspects of white matter architecture and its relationship to EEG/MEG signals (Wolters et al. 2006; Valdez-Hernandez et al. 2009; Gullmar et al. 2010). We were therefore interested in exploring the possibility that signal conduction along white matter tracts may contribute to EEG oscillations in addition to synaptic current dipoles. We suggest a model, meant to serve as a first approximation, that is based on early conductivity models in axons (Hodgkin & Rushton 1946).

In brief, this model is based on two guiding principles. First, because information exchange incurs at a cost that scales with distance, we would expect regions that need to interact at higher bandwidth, higher frequency, and shorter latency to be more proximally located in the brain (Chklovskii & Koulakov 2004; Kriegeskorte & Douglas 2018). From this perspective, shorter path lengths are expected to enable higher frequency communication. However, the majority of short distance connections in the brain are within the gray matter, and a large portion of axons in the neuropil are unmyelinated (Schuz and Braitenberg, 2002). Secondly, transduction along axons is expected to be fast (~2 msec), and therefore unlikely to overlap sufficiently in time to superimpose and produce signals of sufficient strength to be measured at the scalp (Buszaki et al. 2016). However, myelination clearly plays a key role in axonal conduction velocities, and myelination quantification studies (Braitenberg, 1978; Schuz and Braitenberg, 2002; Waxman & Swadlow 1977; Liewald et al. 2014), have demonstrated that even long distance cross fissural fascicles (e.g., corpus callosum) contain a significant portion of unmyelinated fibers (Liewald et al. 2014). Therefore, approximate synchrony of axonal conduction may be possible in unmyelinated neurons, thus providing the opportunity for dipoles to superimpose. Inspired by recent models of "axonal pushing" and the geometry of 'U' fibers (Nie et al. 2012), we explore the additional possibility that transduction from sulcal regions and across fissural (e.g., inter-hemispheric fissure) regions may also contribute to signal.

It should be noted that simply because aspects of the white matter architecture may be used to predict aspects of the EEG signal (e.g., power law distributions), its does not necessarily follow that this is what we are in fact measuring with EEG. Further empirical work is necessary to assess its validity.

This paper is organized as follows. First, we review neurobiologically plausible models that have shaped the central theories of EEG signals and their measurement. We then discuss alternative theories to this central viewpoint, which suggest that additional biological processes (e.g., Fox & O'Brien 1965) or electrodynamics (e.g., corticothalamic) may contribute to EEG oscillations (e.g., Robinson et al. 2001). We then explore the possibility that aspects of white matter architecture are related to EEG biorhythms via simulation. Lastly, we discuss implications of the model for EEG source imaging and BCI implementation

## II. CENTRAL PRINCIPLES OF EEG MEASUREMENT

### A. Geometry of Measurement

EEG measures instantaneous voltage differences at the scalp with respect to a reference or ground. These measurements are thought to reflect the superposition of electric dipoles, which arise when charges of opposite sign are separated by a distance. Dipoles that arise due to synaptic currents in large (e.g., pyramidal) neurons arranged approximately in parallel within the cortical sheet are thought to contribute most significantly to EEG measurements (Adrian & Yamagiwa 1935; Eccles 1951). The geometric consequences of the measurement itself plays a key role in the putative generators of EEG signals.

In the simple case of two opposite charges, $q$, separated by distance, $d$, measured from at distance of $r$, where the measurement distance is significantly larger than the charge separation (ie., r>>d), an approximation of the electric potential is given by:

$$V = \frac{kp\cos\theta}{r^2} \quad (1)$$

Where the dipole moment p is,

$$\vec{p} = q\vec{d} \quad (2)$$

$k$ is Coulomb's constant, and theta is the angle between $d$ and $r$ (Scherg 1990). Within this framework, radial dipoles, oriented approximately orthogonal to the scalp would contribute more significantly to the EEG measurement than tangential dipoles in sulcal regions (See Figure 1(a-c)). Given the geometry of measurement, ensemble and temporally overlapping currents must produce dipoles oriented approximately orthogonal to the scalp.

### B. Temporal Summation and Currents

Neuronal activity in the brain causes transmembrane currents that can be measured in the extracellular medium. Although any type of transmembrane current (e.g., axonal) leads to deflections in intracellular and extracellular field potentials, synaptic transmembrane currents are considered to be the most prominent source of extracellular current flow and local field potentials (LFPs) (Buzsáki et al. 2016). From this perspective, an estimated 10^7 synapses must be approximately synchronously active to generate electric signals sufficient to propagate through the brain, dura, skull, and scalp to be measured by EEG (Cooper et al. 1965).

Although action potentials generate large voltage deflections, they have generally been thought to contribute little to LFPs One of the primary reasons for this is that transduction along the axon is expected to be fast (~2 msec), and therefore unlikely to overlap sufficiently in time to produce signals of sufficient strength to be measured at the scalp (Anderson et al. 1971). Recent evidence shows that numerous non-synaptic processes including synchronous action potentials, spikes, and even ephaptic coupling may contribute to high frequency components of LFPs (Buzsaki et al. 2016). When spike after-hyperpolarizations are also considered, action potentials may also contribute to lower-lower-frequency ranges as well (Einvoll et al. 2007). For example, spiking neurons that generate rhythmic extracellular currents can be phase-locked to lower-frequency oscillations (e.g., Logothetis & Wandell 2004). EEG power in the gamma range and phase of delta band activity appear to be linked to multi-unit activity (Whittingstall & Logothetis 2009). However, the relationship between the full EEG spectral density and LFP activity remains unclear.

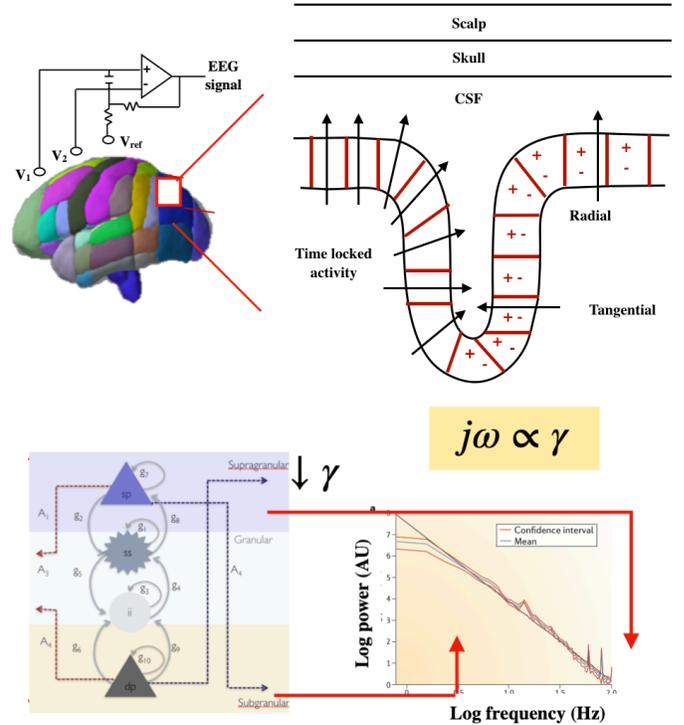

Figure 1: Classic view of EEG neuronal generators. (a) EEG signals are generated by synchronous activity in cortical gray matter, primarily in gyri. (b) Approximately time-locked activity causes summation from synaptic currents across numerous cells in gray matter, thus creating spatially extended radial and tangential dipole moments, inspired by Scherg 1990 (c) canonical microcircuits throughout the cortex generate laminar specific frequency asymmetries (Bastos et al. 2012), whereby (d) superficial laminar layers fire at higher frequency and deeper layers oscillate at lower frequency (Bosman et al. 2012).

### C. Generative Models of EEG

Numerous biophysically plausible models have been proposed for the generation of EEG signals (David & Friston, 2003; Izhikevich & Edelman 2008; Bastos et al. 2012; Moran et al 2013). These may be classified according to neuro-anatomic scale and structure as: cortical, thalamic, or cortical-thalamic (Valdez-Hernandez et al. 2010). In this section, we focus on cortical models, as these have formed the basis for many inverse methods. We consider alternative approaches in subsequent sections.

The observation that archetypal neuronal circuits exist in repeated motifs throughout the brain (Creutzfeldt 1977), gave rise to a number of population density models that attempt to model dynamics within a representative cortical column. In this approach, populations of neurons are modeled as a single "cell" that represents ensemble activity from a specific class of neurons, sometimes referred to as the mean-field approximation (David & Friston 2003). Model structures (i.e., nodes and connectivity) are based on data ranging from neuronal morphology to stereotypic synaptic connections. Their dynamic interactions are typically formulated with coupled differential equations, and can be embedded observationally within a state-space model (e.g., DCM) (Garrido et al. 2007; Moran et al. 2007) )or used for generative purposes (Roebroeck et al. 2011).

Initially, population density models had simple, deterministic architectures and utilized mean activation to approximate population dynamics as a point source (i.e., neural mass models (NMM)) (e.g., Beurle et al. 1956; Griffith et al. 1963; Griffith et al. 1965; Jansen & Rit 1995). Wilson and Cowan created a two compartment model (Figure 2(a)), which was later expanded by Douglas and Martin (1991) to imbue cellular typologies with laminar specificity (Figure 2(b)) (Douglas 1989) . Even in these early models, the critical importance of tightly coupled excitation and inhibition was apparent. Ongoing experimental and theoretical work since have only further demonstrated that energy (ATP) consumption is lower (Sengupta, B. & Stemmler 2014)and coding efficiency is higher in neural circuits with balanced excitation/inhibition (Denève & Machens 2016).

Neural field models (NFM), which have been developed in parallel with NMMs, explicitly incorporate a spatial dimension (Deco et al. 2007), by modeling how neuronal activity unfolds as continuous process along the cortical sheet (Pinotsis et al. 2012). Recent empirical evidence suggests that feedforward connections that link early areas to higher areas (e.g., LGN to V1), originate predominantly in superficial layers and use relatively high frequencies compared to feedback connections (Bosman et al. 2012) (Figure 1(d)). The canonical microcircuit (CMC) model, which has neural field properties, takes these laminar-specific spectral asymmetries into account (Bastos et al. 2012) (Figure 2(c-d)). Recently, the CMC model has been used as a measurement model within DCM to study excitation/inhibition imbalances during seizures (Cooray et al. 2015; Cooray et al. 2015; Papadopoulou et al. 2015).

III. WHITE MATTER ARCHITECTURE AND EEG

Recent studies have investigated the relationship or lack thereof between aspects of the white matter microstructure and EEG signals (Wolters et al. 2006; Valdez-Hernandez et al. 2009; Gullmar et al. 2010). Here, we are interested in understanding the relationship between quantitative aspects of white matter neuroanatomy and its dynamics. Fiber calibre and myelination play an important role in estimating the conduction times, which in turn may provide information about the potential for dipoles to superimpose temporally. We therefore review fiber length, diameter, myelination, and conduction times in axons composing white matter.

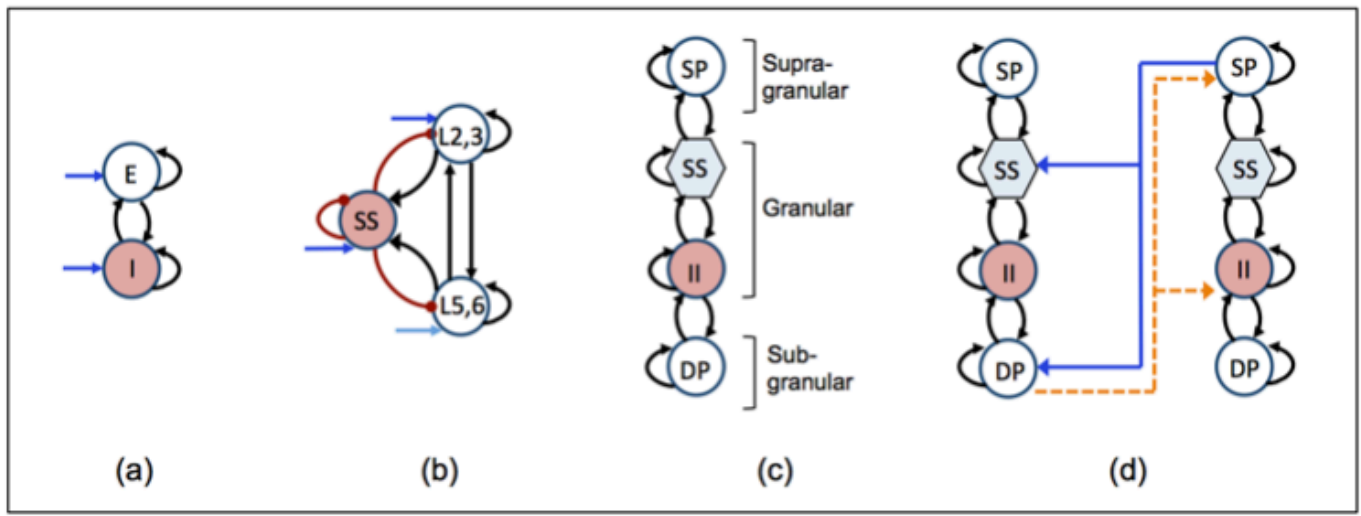

Figure 2. a) Wilson and Cowan (1972) modeled the proportion of excitatory (E) and inhibitory (I) cells firing per unit time via a sigmoidal subpopulation response function. Blue arrows indicate model inputs. (b) The Douglas and Martin (1991) neural mass model comprises balanced excitation (black connections) and inhibition (red connections), adding laminar specificity for excitatory pyramidal cell populations in layers 2,3 (L2,3) and 5,6 (L5,6), and smooth inhibitory cells (SS). Feedforward connections from the thalamus target all populations with less emphasis on L5,6, (blue arrows). (c) The canonical microcircuit model contains four neural populations: superficial pyramidal (SP), spiny stellate (SS), inhibitory interneurons (II), and deep pyramidal (DP) cells across different cortical layers. (d) When the canonical microcircuit is connected into a network, feedforward connections (solid) originate from superficial layers, while feedback (dashed) connections emanate from deeper laminar layers.

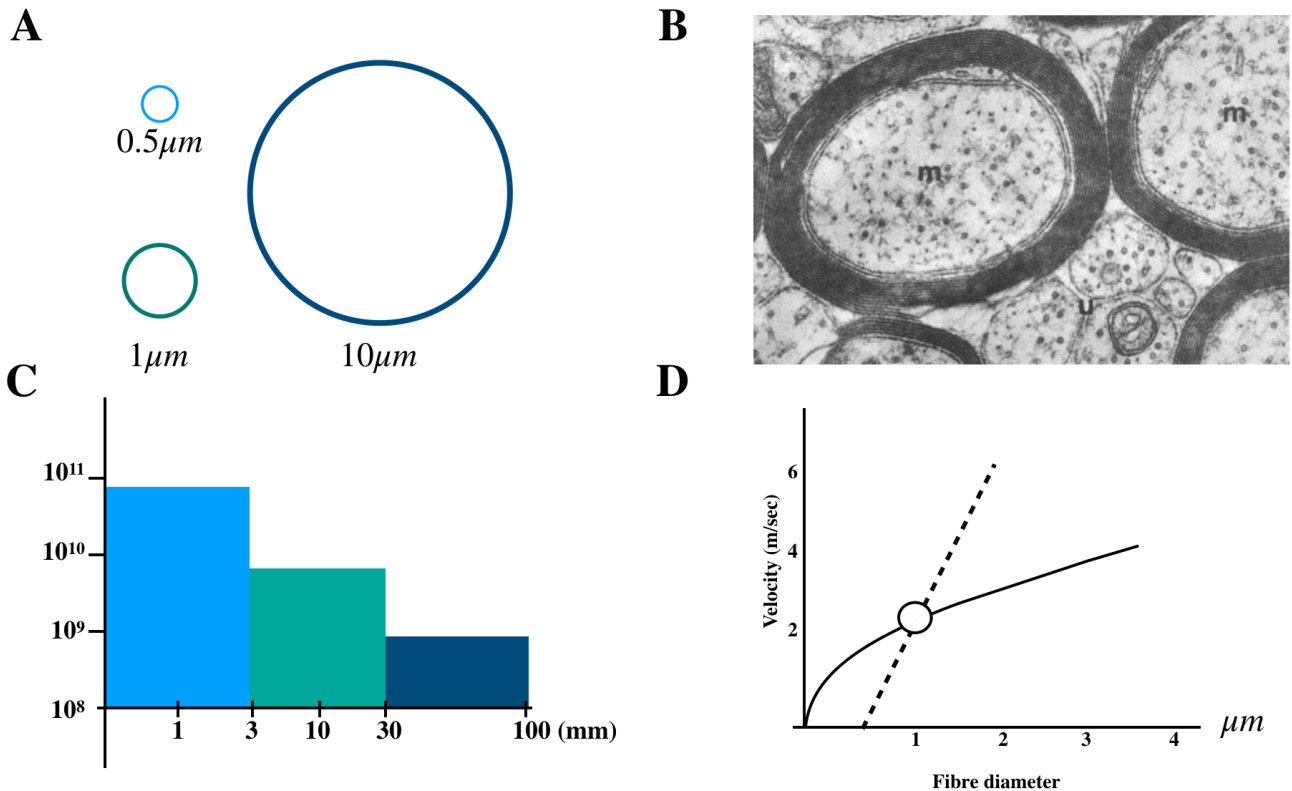

Figure 3. A) Typical diameters of unmyelinated (U), myelinated (M), and large myelinated (LM) neurons, inspired by (Wang et al. 2008). B) Electron micrograph showing myelinated (m) and unmyelinated (u) axons in rabbit corpus callosum, from Waxman & Swadlow, 1977, C) Relative abundance of axons according to length, recreated from Schuz & Braitenberg (2002) D) Predicted relationship between conduction velocity, diameter, and myelination. Small unmyelinated fibers (solid line) conducts faster below 1 micron than myelinated neurons (dashed), figure modified from Rushton (1951)

*A. Fiber Calibre & Myelination*

White matter connections are frequently categorized by length (Thatcher et al. 1986). Short distance (1-3mm) fibers are mostly thin and unmyelinated, connecting nearby cells within the neuropil of gray matter. Intermediate length fibers (10-30mm) connect nearby cortical regions through a system of 'U' shaped fibers. The 'U' fiber system has an interesting geometry. The majority of the "U fibers" connect gyral regions, and the density of axonal fiber endings is significantly higher in gyral as opposed to sulcal regions (Nie et al. 2012). This geometry has led to debate over whether axonal 'pushing' or 'pulling' is responsible for inducing cortical folding during neurodevelopment in gyrencephalic brains (e.g., Nie et al. 2012). Long distance (30-170mm) fascicles connect distant regions of the brain and are located deeper than, or below the 'U' fiber system. The relative abundance of these axon types is shown in Figure 3, and inspired by (Braitenberg, 1978; Schuz and Braitenberg, 2002; Wang et al. 2008) (Figure 3a,b).

There is clearly a tradeoff between processing speed, axonal diameter, and myelination (e.g., Waxman and Swallow 1977). Therefore, it is surprising that relatively few studies have quantified the density of fibers in fascicles, conduction delays, myelination, and axonal diameters in human. This is at least partly because the process of immersing a postmortem brain in fixative can induce tissue damage, particularly in unmyelinated neurons (e.g., Leiwald et al. 2014). Nonetheless, a number of studies have demonstrated that unmyelinated axons appear to be present in significant quantities across a number of species(Wang et al. 2008), even in fascicles that connect distal regions of the brain (e.g., corpus callous) (Caminiti et al. 2013). A histological example of this is shown in (Figure 3c). It is also worth noting that unmyelinated axons are predicted to have faster transduction velocities, when the diameter is smaller than approximately $1\mu m$ (Figure 3d). Therefore, it may be more efficient to use unmyelinated neurons for short distance communication within the gray matter.

*B. Heterogeneous Myelination and Conduction Speeds*

Given that the white matter microstructure is likely a heterogeneous fiber bundle, containing myelinated and unmyelinated neurons with varying axonal diameter, we would expect a diversity of conduction delays. To calculate how potentials spread along white matter tracts, we build on work modeling conduction along cables for axons (Hodgkin and Rushton 1946) and dendritic trees (Rall 1962). When applying nonlinear cable theory to axons, a common assumption was to assume that myelin was perfectly insulating (e.g., McNeal 1976). However, this assumption typically leads to an underestimate of the time constant, and an overestimate of the space constant (Basser et al. 1993). Therefore we consider the contribution of nodes of Ranvier in estimating propagation speeds. To do so, we first calculate spatio-temporal constants for each of the three axonal types.

For myelinated axons of small and large type, we will use the formulation from Andrietti & Bernardini (1984) for a composite cable. Within this formulation, the distribution of potential, $V(x,t)$ may be calculated by the composite equation:

$$\frac{\partial^2 V}{\partial x^x} = (1-\chi)\left(\frac{\tau_m}{\lambda_m^2}\frac{\partial V}{\partial t} + \frac{V}{\lambda_m^2}\right)$$
$$+\chi(x)\left(\frac{\tau_n}{\lambda_n^2}\frac{\partial V}{\partial t} + \frac{V}{\lambda_n^2} + \phi_{ion}(V,t)\right) \quad (3)$$

Where the first term represents the myelinated portion of the axon, and the second represents potential spread in nodal regions. The term $\chi$ selects the contribution of each term along the distance in space (x), and $\tau_m$, $\tau_n$ and $\lambda_m, \lambda_n$ are the time and space constants for myelinated and nodal regions, respectfully. To combine the effects of nodal and myelinated regions, we use the homogenized formulation from Basser et al. 1993 to calculate space and time constants as:

$$\lambda = \left(\left(1-\frac{\delta}{L}\right)\frac{r_a}{r_m} + \frac{\delta}{L}\frac{r_a}{r_n}\right)^{-\frac{1}{2}} \quad (4)$$

And

$$\tau = \lambda^2\left(\left(1-\frac{\delta}{L}\right)r_a c_m + \frac{\delta}{L}r_a c_n\right) \quad (5)$$

Where $r_m$ is the resistance of myelinated membrane (kΩ cm), $r_n$ is the nodal resistance (kΩ cm), $r_a$ is the resistance to axial current flow per unit length of membrane cylinder within the axoplasm (kΩ/cm), $\delta$ is the nodal width (~1x10$^{-4}$ cm), L is the length between nodes. Here, we use the relationship from Hursh 1939, whereby inner axonal diameters are 70% of the ouster diameter in myelinated axons. The membrane capacitance per unit length $c_m$ in (μF/cm), can be calculated by:

$$c_m = \pi d C \quad (6)$$

Here, d is the axonal diameter, and C is the capacitance for a unit area of membrane, which is typically approximated as ~5 μF/cm$^2$ for nodal regions and 5x10$^{-3}$ μF/cm$^2$ for myelinated regions (Tasaki 1955).

For unmyelinated neurons, we use the classic nonlinear cable equations for unmyelinated neurons. It is possible that the bundling of unmyelinated neurons within a fascicle containing myelinated neurons could create an insulating effect. For simplicity, we assume that this insulating effect is negligible. The time and space constants for small unmyelinated (u) axons can be calculated as:

$$\tau_u = r_u c_u \quad (7)$$

$$\lambda_u = \sqrt{\frac{r_u}{r_a}} \quad (8)$$

We assumed that extracellular resistance to longitudinal current flow is negligible, as done typically in cable modeling. In this case, the space constant may be rewritten in terms of resistivity:

$$\lambda_u = \sqrt{\frac{R_u a}{2R_a}} \quad (9)$$

where $R_u$ is the resistance per unit area of an unmyelinated membrane, which is typically ~10$^4$ (Ω/cm$^2$), and $R_a$ is the resistivity of the axoplasm ~140 Ω cm (Schwartz et al. 1979).

Clearly, the magnitude of axon potentials decay with distance, and therefore there is no fixed wave propagating along an action potential. However, an approximation to the speed, or velocity, of electronic conduction can be calculated as (Jack et al. 1975):

$$\nu = 2\frac{\lambda}{\tau} \quad (10)$$

Here, we converted to units of m/sec to make comparisons with conductions reported in the literature. Using these equations and parameters, we estimated spatio-temporal constants and conduction speeds for each neuronal subtype (see Table 1).

|  | Diameter (μm) | $\tau$ (msec) | $\lambda$ (cm) | $\nu$ (m/sec) |
|---|---|---|---|---|
| **Unmyelinated** | 0.5 | 0.1 | 0.00084 | 0.27 |
| **Myelinated** | 1 | 0.192 | 0.0537 | 5.58 |
| **Large Myelinated** | 10 | 0.192 | 0.170 | 17.6 |

These estimated conduction velocities appear reasonable, given reported values in the literature. Waxman & Swadlow (1977) recorded callosal conduction velocities in adult, unanesthetized rabbits from cells bodies of neurons which send an axon across the corpus callosum (callosal efferent neurons). Reported conduction velocities ranged from 0.3 to 12.9 m/sec. Similar conduction velocities from Caminiti et al. (2013) who estimated mean conduction velocities in corpus callosum in humans to range from 7.83-9.81 m/sec. Overall, it is important to note that the large myelinated neurons make up only a small portion of fibers, even in large fascicles (Wang et al. 2008).

*C. Estimating Conduction Along Fascicles*

In considering the potential for conduction along white matter fascicles to contribute to EEG measurements, there are two potential geometric considerations that we suggest would be likely locations for dipoles. First, when white matter fibers cross a fissure (e.g., inter hemispheric fissure), a signal would need only travel through cerebrospinal fluid (CSF) and bone in order to reach the scalp. Volume conduction calculations are also simplified in this case, as there are fewer boundary conditions to consider. Figure 4a shows a coronal stain in Macaca Mulatta that has been stained for myelin, showing the possibility of dipoles arising from the corpus callosum.

Figure 4b shows the potential geometry of dipole sources traveling along the white matter in the 'U' fiber system. Here, if dipoles arise between inside and outside of a fascicle, then radial dipoles will be strongest in the sulcal regions. This might occur if a dipole arises due to charge differences between the axonal cytosol and the extracellular milieu, or if a charge separation exists between the inside and outside of a fascicle. This might occur if a number of unmyelinated neurons, presumably firing quasi-synchronously, were insulated by myelinated neurons.

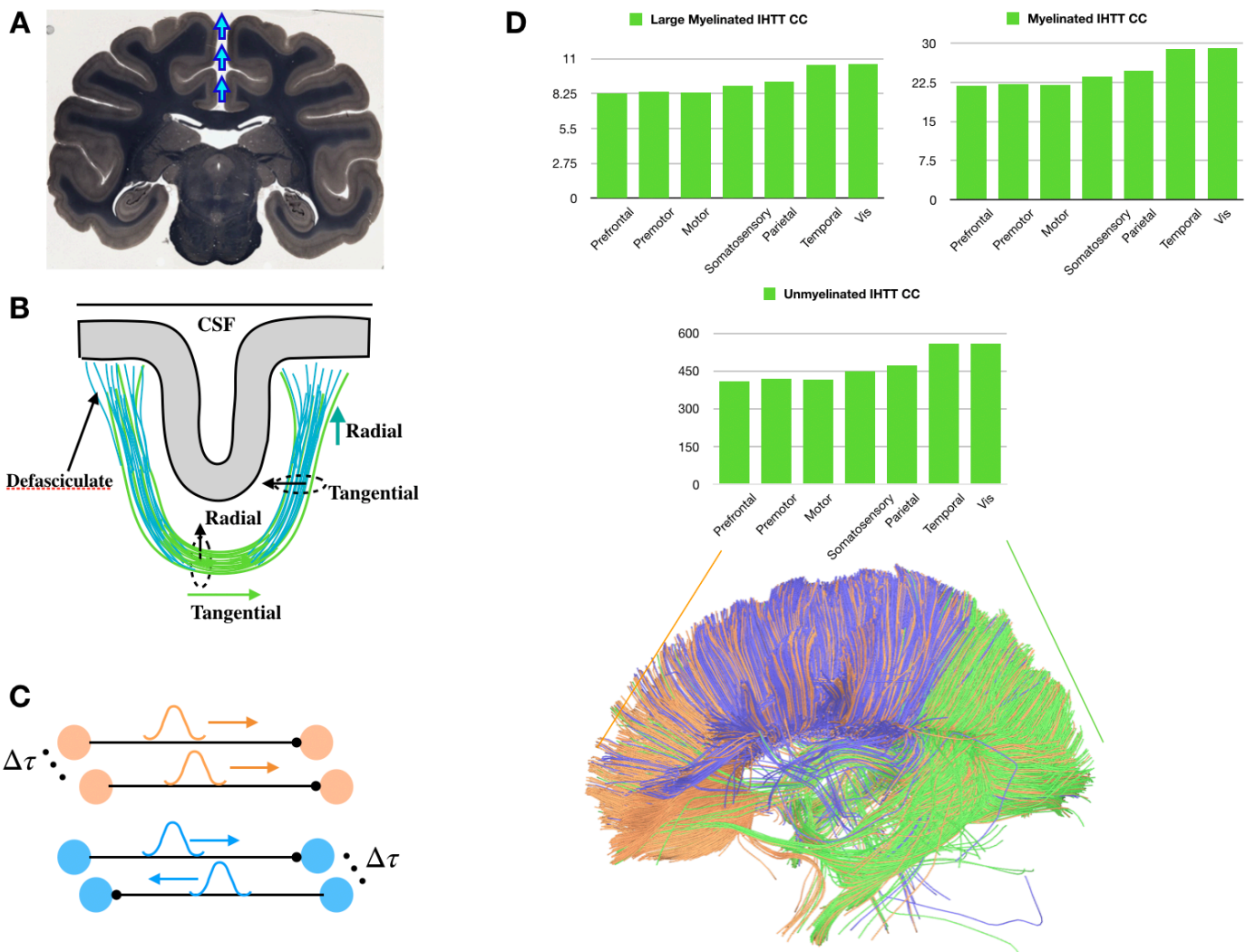

Figure 4. *Potential White Matter Contributors to EEG activity.* A) Coronal section of an adult Macaca Mulatta brain using the Weil stain, showing myelinated regions in dark blue. Image courtesy of the open data base at http://brainmaps.org. Breaches of electrical activity from fissural regions such interhemispheric regions may contribute to EEG activity, shown in light blue (B) Tangential and Radial components shown for a 'U' fiber for potentials along the fasciculus and between the fasciculus and the extracellular milieu. (C) Feedforward and recursive transduction, inducing a small delay due to a refractory period, or synaptic delay in the "ping pong" recursive case, D) Interhemispheric transit time estimates along the Corpus Callosum, shown for large myelinated, myelinated, unmyelinated, and weakly insulated unmyelinated axons. Frequency estimates are also shown for the two unmyelinated cases.

We were interested in using our estimated velocities to calculate transit times along tracts for each of the fiber types. We first calculated interhemispheric transfer time (IHTT) across the CC, as these times have been well studied. Because CC length varies with anterior-posterior connections, we used specific values ranging from 110.4-150.7 reported by Caminiti et al. 2013. For the CC, the interhemispheric transit time ranges for myelinated neurons ranged from 4.5-27.0 msec (Figure 4). This is consistent with findings from Westerhausen et al. (2006) who found inter-hemispheric transfer times ranging from 2.5-18.6 msec, based on subtracting evoked and response times between hemispheres. For unmyelinated neurons, IHTT exceeded 500msec in the visual and occipital part of the CC with a longer length.

Length estimates were difficult to obtain from the literature, apart from the corpus callosum (CC). We therefore derived volumetric and length estimates from diffusion MRI data. In brief, we used methods available in DSI Studio (Yeh et al. 2010), applied to the Human Connectome Project 1021 group average image, and extracted tract parameters based on seeds from the JHU 1mm atlas. Supplementary Table1 contains these extracted parameters.

If conduction along white matter tracts contributes to the biorhythms of the brain, it is unlikely that conduction along large myelinated neurons is the major contributor. The speeds are simply far too rapid, and expected to be above line noise, and therefore not often studied with EEG techniques. We therefore chose to explore the possibility that transduction in unmyelinated neurons was responsible for these rhythms. This possibility seems unlikely, given the spatial decay of signals. However, it also seems unlikely that unmyelinated neurons would represent a significant portion in major fascicles, if not for the purpose of carrying meaningful information.

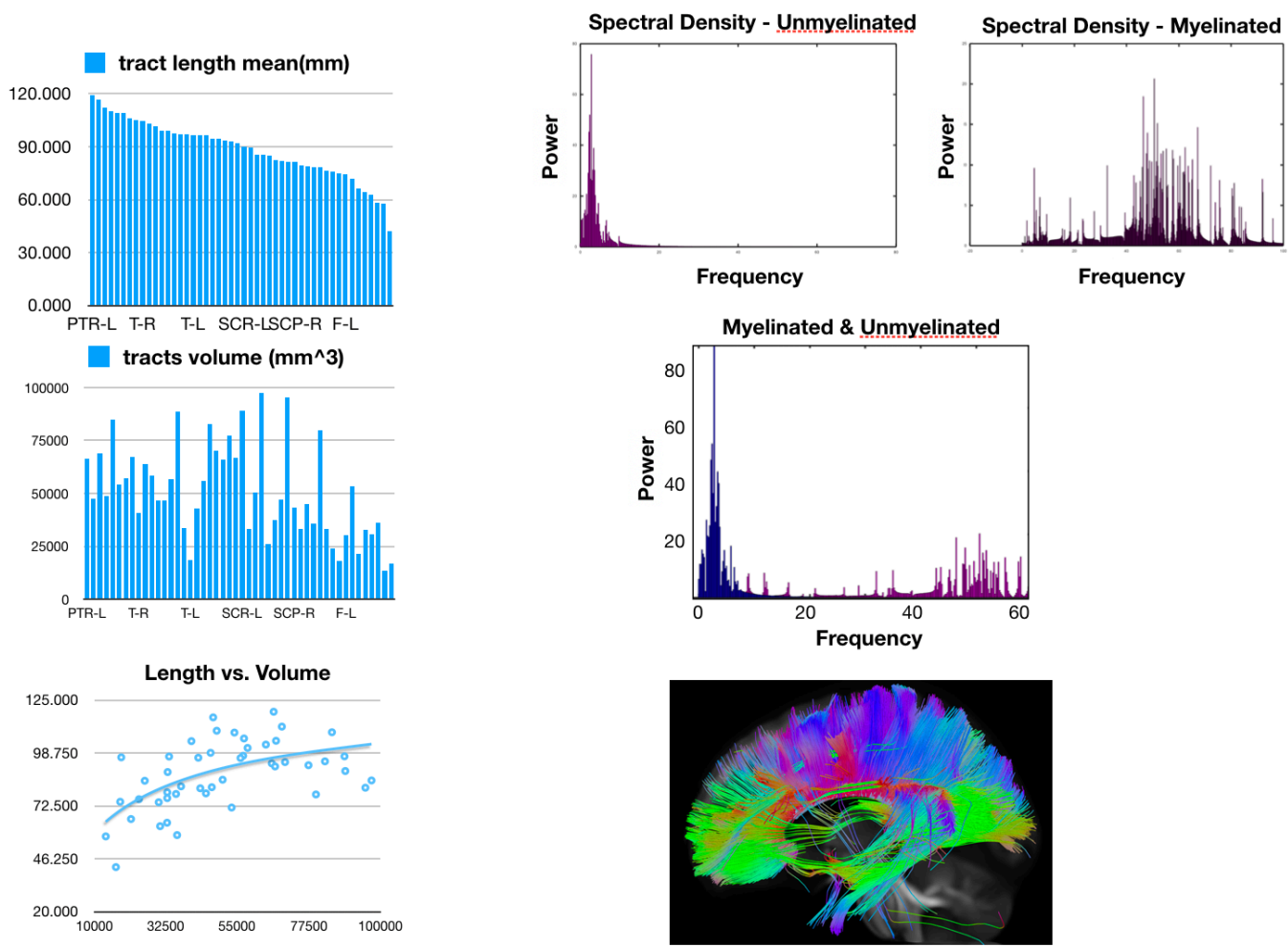

Figure 5. *Tract Lengths, Volumes, and Predicted Frequencies* A) Tract lengths estimated from the Human Connectome Project 1021 group average release of diffusion MRI data. Lengths were estimated using DSI Studio software, shown in decreasing order. B) Corresponding tract volumes (C) Scatter plot of length and value shown with an logarithmic fit. D) Power spectral densities for frequencies estimated from myelinated and E) unmyelinated fibers using volumetric and length estates from the F) diffusion HCP 1021 data.

To approximate frequency, we calculated the inverse of the time required to traverse the length of a given fascicle. We assumed that only a single waveform may occupy a given connection at a time, because a given neuron is expected to undergo a refractory period before firing a subsequent signal. We assume that there would be a small delay in the form of either a refractory period, or in a synaptic delay, before the next 'up' state reaches the same location in the retrograde direction Figure 4c. Refractory periods generally range from 0.6-2.0 msec (e.g., Waxman and Swandel 1977). We therefore selected 1msec, as an intermediate delay.

Figure 5a shows the extracted lengths and volumes for all of the major fascicles as identified by the JHU atlas. The corresponding volumes for these tracts are shown in Figure 5b. Length and volume are plotted together along with a logarithmic curve fit. In Figure 5c. We used the volume to approximate the power of the signal. The estimated power spectral density for unmyelinated and myelinated axons is shown on a log scale Figure 5d.

### D. Considerations for Source Imaging

Cortical models have form the bases for the majority of EEG source methods. Although useful, inverse methods based on local assumptions alone often result in considerable spatial mismatch with other functional neuroimaging modalities (e.g., functional MRI (fMRI)) (Daunizeau et al. 2007) or utilize circular logic where other modalities (e.g., PET, fMRI) inform their location while also being used to confirm it. Few biophysical models of EEG generators explain observed phenomena such as traveling neocortical waves (Zhang et al. 2018).

Should white matter conduction contribute significantly to the EEG signal, both forward and inverse models could be adjusted. Forward models may include sulcal and fissural regions as potential source locations. Early inverse solutions considered only a few focal sources and assumed temporal independence of measurements (e.g., Scherg et al. 1999). However, activity is continuous throughout the brain and often coherent between spatially distinct loci. Therefore, continuous current source models which attempt to determine the strength and orientation of dipoles at all possible locations are more biologically plausible. Methods that consider coherence (e.g., Mahjoory et a. 2017) in light of expected delays, and coincidence detection based structural connectivity distances and frequency (e.g., Lei et al. 2015) may be more useful should white matter contribute to EEG signals.

### IV. DISCUSSION

In the present work, we were concerned with modeling EEG biorhythms using conventional models of electronic conduction in nerve fibers applied to white matter structure. Unsurprisingly, our simulations of conduction in myelinated neurons in the white matter revealed were very rapid, and would theoretically produce frequencies faster than those typically examined with EEG. This finding led us to examine the possibility that unmyelinated fibers that have slower conduction speeds may contribute to EEG rhythms. Cable theory combined with parameters derived from diffusion MRI, yielded estimates that were capable of producing lower frequencies that appeared to approximately follow a 1/f distribution. From the parameters reported here, it can also be shown that a range of frequency estimates could be generated within the laminar layers of the neuropil, depending on connection lengths, which may yield estimates similar to those expected with CMC models. Our model did not include the 'U' fiber system, which may contribute to measurements in intermediate and higher frequencies.

Nonetheless, our model ignored space constants, and it is unlikely that unmyelinated neurons could propagate signals over the distances traversed by the major fascicles. Conversely, it also improbable that the brain would support the energetic demands of these neurons, if they did not serve a purpose. Within a fascicle, it is possible that myelinated neurons provide a form of insulation that would enable signals to be carried over longer distances in small diameter unmyelinated neurons. However, this possibility of insulation would also potentially yield diminished signal transduction times, and thus, higher frequencies. In this case, it is unclear how the lower frequency rhythms (e.g., delta) which present in abundance would be generated from white matter propagation delays

Future studies focusing on propagation via unmyelinated neurons in fascicles remains attractive for a number of reasons. First, in reviewing the neuroanatomy, it is surprising that unmyelinated neurons putatively make up a significant portion of connections in the neuropil and overall white matter volume (Schuz and Braitenberg 2002), yet receive surprisingly little attention. In many species, unmyelinated axons make up ~30% of connections in the CC (Leiwald et al. 2014). Due to brain size, it is expected that this ratio may be slightly lower in humans (e.g., Wang et al. 2008), perhaps ~16%, yet still significant. Should white matter (axon potentials) contribute significantly to EEG signal, at least an order of magnitude less neurons would be required to fire semi synchronously. It is also interesting that fiber endings have the tendency to fan out as they defasciculate (Figure 3). Therefore approximately synchronous signaling would not necessarily need to emanate from or terminate in a spatially contiguous gyral region. This may reduce the total number of spatially extended neurons that must fire synchronously to superimpose to produce an EEG signal. Future studies aimed at quantifying the spatio-temporal overlap required for white matter to yield a measurable EEG signal may shed light into this issue. Additionally, empirical measurements in animal models may provide insight into the extent to which fascicle bundling effects impedance and capacitance, as well as signals propagating to surface EEG via sulcal breeches and fissure crossings.

ACKNOWLEDGMENT

PKD would like to thank the Brain & Behavior Research Foundation, NIH-NIMH K Award, and Renee Hartig at the

Max Planck Tubingen for very helpful conversations about neuroanatomy, and sharing posters from Schuz & Braitenberg.

# Supplementary Table

| Tract Name | Abbreviation | number of tracts | tract length mean(mm) | length sd(mm) | volume (mm^3) |
|---|---|---|---|---|---|
| Body_of_corpus_callosum | CCB | 4780 | 85.2769 | 38.9629 | 97284 |
| Genu_of_corpus_callosum | CCG | 4517 | 71.8266 | 33.6682 | 53259 |
| Splenium_of_corpus_callosum | CCS | 4850 | 109.178 | 53.3539 | 84824 |
| Fornix | FOR | 4503 | 58.165 | 44.762 | 36101 |
| Corticospinal_tract_R | CST-R | 4732 | 96.5472 | 42.9579 | 42773 |
| Corticospinal_tract_L | CST-L | 4824 | 78.8615 | 45.6761 | 45173 |
| Medial_lemniscus_L | ML-L | 4992 | 78.5662 | 43.3368 | 35759 |
| Medial_lemniscus_R | ML-R | 4985 | 89.4362 | 43.7444 | 33126 |
| Inferior_cerebellar_peduncle_L | ICP-L | 4853 | 64.3001 | 40.1232 | 32946 |
| Inferior_cerebellar_peduncle_R | ICP-R | 4870 | 62.5847 | 38.4408 | 30755 |
| Superior_cerebellar_peduncle_R | SCP-R | 4808 | 81.3828 | 45.717 | 43396 |
| Superior_cerebellar_peduncle_L | SCP-L | 4612 | 76.5159 | 46.4511 | 33001 |
| Cerebral_peduncle_R | CP-R | 4795 | 109.024 | 39.4683 | 54133 |
| Cerebral_peduncle_L | CP-L | 4605 | 97.5215 | 47.1697 | 56874 |
| Anterior_Internal_capsule_R | AIC-R | 4595 | 81.9135 | 33.6674 | 47014 |
| Posterior_Internal_capsule_R | PIC-R | 4950 | 103.056 | 34.8885 | 64035 |
| Anterior_Internal_capsule_L | AIC-L | 4850 | 85.6298 | 41.9897 | 50425 |
| Posterior_Internal_capsule_L | PIC-L | 4922 | 104.937 | 39.3593 | 67231 |
| Retrolenticular_Internal_capsule_R | RIC-R | 4923 | 96.375 | 51.7331 | 56061 |
| Tapetum_L | T-L | 2960 | 96.731 | 55.108 | 18522 |
| Tapetum_R | T-R | 4687 | 104.766 | 56.5038 | 40606 |
| Uncinate_fasciculus_L | UF-L | 4796 | 75.9748 | 37.6875 | 24094 |
| Uncinate_fasciculus_R | UF-R | 4702 | 74.6931 | 27.4525 | 18192 |
| Superior_fronto-occipital_fasciculus_L | SFO-L | 4940 | 82.3408 | 47.2307 | 37340 |
| Superior_fronto-occipital_fasciculus_R | SFO-R | 4961 | 79.3827 | 37.1924 | 33001 |
| Superior_longitudinal_fasciculus_L | SLF-L | 4973 | 94.3778 | 38.1929 | 70067 |
| Superior_longitudinal_fasciculus_R | SLF-R | 4958 | 92.2144 | 40.5398 | 66929 |
| Fornix_(cres)_/_Stria_terminalis_R | F-R | 4895 | 66.1348 | 43.6731 | 21596 |
| Fornix_(cres)_/_Stria_terminalis__L | F-L | 4804 | 74.4608 | 38.247 | 30313 |
| Cingulum_(hippocampus)_L | CH-L | 4704 | 57.5224 | 34.0456 | 13657 |
| Cingulum_(hippocampus)_R | CH-R | 4482 | 42.2925 | 29.5971 | 16840 |
| Cingulum_(cingulate_gyrus)_L | CG-L | 4816 | 85.1146 | 38.9967 | 25920 |
| Cingulum_(cingulate_gyrus)_R | CG-R | 4980 | 97.1333 | 38.9933 | 33574 |
| External_capsule_L | EC-L | 4252 | 111.969 | 56.7219 | 69075 |
| External_capsule_R | EC-R | 4940 | 93.7338 | 55.9257 | 65883 |
| Sagittal_stratum_L | SS-L | 4811 | 116.548 | 52.8905 | 47448 |
| Sagittal_stratum_R | SS-R | 4968 | 109.928 | 54.0833 | 48570 |
| Posterior_thalamic_radiation_L | PTR-L | 4849 | 119.348 | 52.9114 | 66482 |
| Posterior_thalamic_radiation__R | PTR-R | 4962 | 106.096 | 53.6125 | 57151 |
| Posterior_corona_radiata_L | PCR-L | 4813 | 97.2365 | 51.3582 | 88765 |
| Posterior_corona_radiata_R | PCR-R | 4944 | 92.8232 | 44.9294 | 77446 |
| Retrolenticular_Internal_capsule_L | RLP-L | 4923 | 101.393 | 42.4317 | 58281 |
| Anterior_corona_radiata_R | ACR-R | 4933 | 78.3297 | 40.9959 | 79795 |
| Anterior_corona_radiata_L | ACR-L | 4922 | 81.6669 | 43.4841 | 95349 |
| Superior_corona_radiata_R | SCR-R | 4968 | 94.7467 | 38.4471 | 82631 |
| Superior_corona_radiata_L | SCR-L | 4827 | 89.9375 | 41.2986 | 89023 |